\documentclass{elsart}
\usepackage{amssymb}
\usepackage[german,english]{babel}
\usepackage[intlimits,sumlimits,namelimits]{amsmath}
\usepackage[dvips]{epsfig}
\usepackage[dvips]{color}
\usepackage{ifthen,shadow}
\usepackage{fancyhdr}

\newcommand{\beq}{\begin{equation}}
\newcommand{\eeq}{\end{equation}}
\newcommand{\beqa}{\begin{eqnarray}}
\newcommand{\eeqa}{\end{eqnarray}}
\newcommand{\bra}[1]{\mbox{$\langle #1|$}}
\newcommand{\ket}[1]{\mbox{$|#1\rangle$}}

\begin{document}

\begin{frontmatter}

\title{\bf Single-spin asymmetries of $d(\gamma,\pi)NN$ in the first 
resonance region}

\author{Eed M.\ Darwish\corauthref{eed}}
\corauth[eed]{Present address: Physics Department, Faculty of Science, 
South Valley University, Sohag, Egypt (e-mail: eeddarwish@yahoo.com).}

\address{Institut f\"{u}r Kernphysik, Johannes Gutenberg-Universit\"{a}t,
       J.J.\ Becher-Weg 45, D-55099 Mainz, Germany}

\date{\today}

\begin{abstract}
Incoherent photoproduction of pions on the deuteron in the first resonance
region is investigated with special emphasis on single-spin asymmetries.
For the elementary pion production operator an effective Lagrangian model
which includes the standard pseudovector Born terms and a resonance
contribution from the $\Delta$(1232)-excitation is used. Single-spin
asymmetries, both for charged and neutral pion photoproduction on the
deuteron, are analyzed and calculated in the first resonance region. 
The linear photon asymmetry $\Sigma$, vector target
asymmetry $T_{11}$ and tensor target asymmetries $T_{20}$, $T_{21}$, and
$T_{22}$ for the reaction $d(\gamma,\pi)NN$ with polarized photon beam and/or 
oriented deuteron target are predicted for forthcoming experiments.

\vspace{0.2cm}

\noindent{\it PACS:}
24.70.+s; 13.60.Le; 25.20.Lj\\
\noindent{\it Keywords:}  Polarization phenomena in reactions; Meson 
production; Photoproduction reactions.
\end{abstract}
\end{frontmatter}


\section{Introduction} 
\label{sec1}
It has been known for a long time that the electromagnetic probe belongs to 
the important tools in investigating the structure and properties of the 
strongly interacting particles and nuclei. It is not only because its 
properties are well known but also because it is weak enough, so one can
treat it in any reaction perturbatively. Meson production is the primary
absorptive process on the nucleon. It proceeds mainly through the intermediate
excitation of a nucleon resonance and gives important information on the
internal nucleon structure. Therefore, it provides stringent tests of any kind
of hadron models.

During the last years, photo- and electroproduction of pions on a single
proton have throughly been studied both theoretically and experimentally.
Since the deuteron is the simplest nucleus containing a neutron, the process
of pion production on the deuteron can be used for examining pion production
on a neutron. It can also gives information on pion production on off-shell
nucleons, as well as on the very important $\Delta N$-interaction in a
nuclear medium.

The study of polarization observables in pseudoscalar meson production in
electromagnetic reactions on light nuclei has become a very active
field of research in medium-energy nuclear physics with respect to the
investigation of small but interesting dynamical effects, which normally are 
buried under the dominant amplitudes in unpolarized total and differential 
cross sections, but which often may show up significantly in certain 
polarization observables. The reason for this feature lies in the fact that 
such small amplitudes or small contributions to large amplitudes may be 
amplified by interference with dominant amplitudes, or that dominant 
amplitudes interfere destructively leaving thus more room to the small 
amplitudes. Polarization observables will also give additional valuable 
information for checking the spin degrees of freedom of the elementary pion 
production amplitude of the neutron, provided, and this is very important, 
that one has under control all interfering interaction effects which prevent 
a simple extraction of this amplitude.

Recent work in view of continuing technical improvements, such as
ELSA in Bonn, LEGS in Brookhaven, CEBAF in Newport News, or MAMI
in Mainz, for preparing polarized beams and targets and for
polarimeters for the polarization analysis of ejected particles,
has motivated the examination of certain aspects of the
polarization observables in pion production on the deuteron that
are fundamental to the process. Quasifree $\pi^-$ photoproduction
on the deuteron via the $\gamma d\to\pi^-pp$ reaction has been 
investigated within a diagrammatic approach including $NN$- and 
$\pi N$-rescattering effects~\cite{Log00}. In that work, the authors 
reported predictions for the squared moduli of amplitudes 
$\mid T_{fi}\mid^2$, analyzing powers connected to beam polarization 
$T_{22,00}$, to target polarization $T_{00,20}$, and to polarization 
of one of the final protons $P1_y$. It has been shown, that final state 
interaction effects play a noticeable role in the behaviour of these 
observables. In our previous evaluation~\cite{Dar03,Dar03+}, the energy 
dependence of the three charge states of the pion for the 
$\gamma d\rightarrow\pi NN$
reaction over the whole $\Delta$(1232)-resonance region has been
evaluated. We have presented results for differential and
total cross sections as well as the spin asymmetry and the 
Gerasimov-Drell-Hearn (GDH) sum rule for the deuteron. 

Notwithstanding this continuing effort to study this process, the
wealth of information contained in it has not yet been fully
exploited. Since the $t$-matrix has 12 independent complex
amplitudes and in order to determine completely the $t$-matrix,
one has to measure 23 independent observables. Up to present
times, only a few observables have been measured and studied in
details, e.g., differential and total cross sections. Therefore,
in~\cite{Dar0309} we have investigated incoherent single pion
photoproduction on the deuteron in the $\Delta$(1232)-resonance
region with special emphasis on spin-polarization observables. In
that work, we have presented some results for the $\pi$-meson
spectrum with other spin-observables as functions of pion momentum
at different values of pion angles for photon energy only at the
$\Delta$(1232)-resonance region.

Our main aim in this contribution is to investigate incoherent
single pion photoproduction on the deuteron in the first resonance 
region with special emphasis on single-spin asymmetries. 
Particularly, the scope of this article is to predict some
additional polarization observables which are experimentally
measured or planned to be measured at different laboratories, for
instance, the linear photon asymmetry $\Sigma$ for the reaction
$\vec\gamma d\to\pi^0np$ which is measured\footnote{Data are still
preliminary.} most recently at LEGS Brookhaven National 
Laboratory~\cite{Lucas}. Furthermore, results for all the three
isospin channels of the reaction $d(\gamma,\pi)NN$ with polarized
photon beam and/or oriented deuteron target at different photon
lab-energies will be reported. The importance of this process
comes from the fact that the deuteron, being the simplest nuclear
system, plays a similar fundamental role in nuclear physics as the
hydrogen atom plays in atomic physics. 

The paper is organized as follows. In Section~\ref{sec2} the elementary
pion production operator on the free nucleon which we use as input in
the calculations on the deuteron is presented. In Section~\ref{sec3}
we outline the formalism of incoherent single pion photoproduction on the
deuteron. The general form of the differential cross section is also
introduced in this section. The treatment of the $\gamma d\to\pi NN$
transition matrix elements, based on time-ordered perturbation theory, is
described in Section~\ref{sec4}. Section~\ref{sec5} is devoted to the
central topic of this paper. The complete formal expressions of
single-spin asymmetries of the reaction $\gamma d\to\pi NN$ with polarized 
photon beam and/or oriented deuteron target in terms of the transition matrix 
amplitudes are developed in this section. Details of the actual calculation
and the results are presented and discussed in Section~\ref{sec6}. Finally,
we conclude and summarize our results in Section~\ref{sec7}.

\section{Elementary process}
\label{sec2}
The most important ingredient of our model is the elementary operator for
pion photoproduction on a single nucleon which is the starting point of
the construction of an operator for pion photoproduction in the two-nucleon
space. In this work we will examine the various observables for pion photoproduction
on the free nucleon using, as in our previous work~\cite{Dar03,Dar03+,Dar0309}, the
effective Lagrangian model developed by Schmidt {\it et al.}~\cite{ScA96}. The
main advantage of this model is that it has been constructed to give a
realistic description of the $\Delta$(1232)-resonance region.  It is also
given in an arbitrary frame of reference and allows a well defined off-shell
continuation as required for studying pion production on nuclei. This model 
contains besides the standard pseudovector Born terms a resonance contribution
from the $\Delta$(1232)-excitation. For details with respect to the
elementary pion photoproduction operator we refer to~\cite{ScA96}.
As shown in Figs.~1, 2, and 3 in our previous work~\cite{Dar03}, the results
of our calculations for the elementary process are in good agreement with
recent experimental data as well as with other theoretical predictions and
gave a clear indication that this elementary operator is quite satisfactory
for our purpose, namely to incorporate it into the reaction on the deuteron.

\section{Basic formalism}
\label{sec3}
In this section we present the formalism of incoherent single pion 
photoproduction on the deuteron
\beqa
\gamma(k) + d(d) &\to& \pi(q)+ N_1 (p_1) + N_2(p_2)
\eeqa
where $k=(\omega_\gamma,\vec k\,)$ and $d=(E_d,\vec d\,)$ denote the initial 
photon and deuteron four-momenta, respectively. The four-momenta of final 
meson and two nucleons are denoted by $q=(\omega_q,\vec q\,)$ with 
$\omega_{q} = \sqrt{m_{\pi}^{2} + \vec{q}^{\,2}}$, $m_{\pi}$ as pion mass,
and $p_j=(E_j,\vec p_j\,)$ ($j=1,2$) with $E_{j}=\sqrt{M_{N}^{2} + 
\vec{p}_{j}^{\,2}}$, respectively, and $M_N$ as nucleon mass.

The general expression of the cross section is given, using the conventions
of Bjorken and Drell~\cite{BjD64}, by
\beqa
d\sigma &=& (2\pi)^{4}\delta^{4}\left( k+d-q-p_{1}-p_{2}\right) 
\frac{1}{|\vec{v}_{\gamma}-\vec{v}_{d}|} \frac{1}{2\omega_{\gamma}} 
\frac{1}{2E_{d}} \frac{d^{3}q}{(2\pi)^{3}} \frac{1}{2\omega_{q}} 
\nonumber \\ 
& & \times
~ \frac{1}{2} \frac{d^{3}p_{1}}{(2\pi)^{3}}\frac{M_N}{E_{1}}    
\frac{d^{3}p_{2}}{(2\pi)^{3}}\frac{M_N}{E_{2}} 
~ \frac{1}{6}\sum_{smtm_{\gamma}m_d} 
|{\mathcal M}^{(t\mu)}_{s m m_{\gamma} m_d}(\vec{k},\vec{d},\vec{q},
\vec{p_1},\vec{p_2})|^{2} \, , 
\label{diff}
\eeqa
where $m_{\gamma}$ denotes the photon polarization, $m_{d}$ the spin 
projection of the deuteron, $s$ and $m$ total spin and projection of the 
two outgoing nucleons, respectively, $t$ their total isospin, $\mu$ the 
isospin projection of the pion, and $\vec{v}_{\gamma}$ and $\vec{v}_{d}$ the 
velocities of photon and deuteron, respectively. The states of all particles 
are covariantly normalized. The reaction amplitude is denoted by 
${\mathcal M}^{(t\mu)}_{sm m_{\gamma}m_d}$. As in~\cite{ScA96}, we have chosen 
as independent variables the pion momentum $q$, its angles $\theta_{\pi}$ and
$\phi_{\pi}$, the polar angle $\theta_{p_{NN}}$ and the azimuthal
angle $\phi_{p_{NN}}$ of the relative momentum $\vec p_{NN}$ of the
two outgoing nucleons.

The total and relative momenta of the final $NN$-system are defined by 
$\vec{P}_{NN} = \vec{p}_{1} + \vec{p}_{2}= \vec{k} - \vec{q}$ 
and $\vec p_{NN} = \frac{1}{2}\left(\vec{p}_{1} - \vec{p}_{2}\right)$, 
respectively. The absolute value of the relative momentum $\vec p_{NN}$ 
is given by 
\beqa
\label{relm}
p_{NN} = \frac{1}{2}\sqrt{ \frac{ E_{NN}^{2}(W_{NN}^{2}-4
M_{N}^{2}) }{E_{NN}^{2}-P^{2}_{NN}\cos^{2}\theta_{Pp_{NN}}} }\, ,
\eeqa
where $\theta_{Pp_{NN}}$ is the angle between $\vec{P}_{NN}$ and 
$\vec p_{NN}$. $E_{NN}$ and $W_{NN}$ denote total energy and invariant mass of 
the $NN$-subsystem, respectively.

For the evaluation we have chosen the laboratory frame where
$d^{\mu}=(M_d,\vec 0\,)$ with $M_d$ as deuteron mass. As coordinate
system a right-handed one is taken with $z$-axis along the
momentum $\vec k$ of the incoming photon and $y$-axis along 
$\vec k\times\vec q$. Thus, the outgoing pion defines the
scattering plane. Another plane is defined by the momenta of the 
outgoing nucleons which we will call the nucleon plane 
(see Fig.~4 in~\cite{Dar03}).

Finally, we get the differential cross section of incoherent single pion 
photoproduction on the deuteron as
\beqa
\frac{d^2\sigma}{d\Omega_{\pi}} &=& \int_{0}^{q_{\rm max}}dq
~\int d\Omega_{p_{NN}} ~\mathcal{K} ~\frac{1}{6}~
\sum_{smtm_{\gamma}m_d} |{\mathcal M}^{(t\mu)}_{sm m_{\gamma}m_d}(\vec{k},
\vec{q},\vec{p_1},\vec{p_2})|^{2}\,,
\label{fivefold}
\eeqa
where the phase space factor $\mathcal{K}$ is expressed in terms of relative and 
total momenta of the two final nucleons as follows
\beqa
\label{rhos}
\mathcal{K} & = & \frac{1}{(2\pi)^{5}}\frac{p_{NN}^{2}M_{N}^{2}}
{\left| E_{2} (p_{NN}+\frac{1}{2} P_{NN} 
\cos\theta_{Pp_{NN}}) + E_{1}
(p_{NN}-\frac{1}{2} P_{NN} \cos\theta_{Pp_{NN}}) \right| } 
\nonumber \\ & & \times
~\frac{q^{2}}{16\omega_{\gamma}M_{d}\omega_{q}} \, .
\eeqa

\section{The transition $\mathcal M$-matrix}
\label{sec4}
The general form of the photoproduction transition matrix is given by
\beqa
{\mathcal M}^{(t\mu)}_{sm m_{\gamma}m_d}(\vec{k},\vec{q},\vec{p_1},\vec{p_2})
& = & ^{(-)}\bra{\vec{q}\,\mu,\vec{p_1}\vec{p_2}\,s\,m\,t-\mu}\epsilon_{\mu}
(m_{\gamma})J^{\mu}(0)\ket{\vec{d}\,m_d\,00}\, , 
\eeqa
where $J^{\mu}(0)$ denotes the current operator and
$\epsilon_{\mu}(m_{\gamma})$ the photon polarization vector. The
electromagnetic interaction consists of the elementary production
process on one of the nucleons $T_{\pi\gamma}^{(j)}$ $(j=1,2)$ and in
principle a possible irreducible two-body production operator
$T_{\pi\gamma}^{(NN)}$. The final $\pi NN$ state is then subject to
the various hadronic two-body interactions as described by an
half-off-shell three-body scattering amplitude $T^{\pi NN}$. In principal, 
one must take into account all of the possible terms in the calculation 
of the transition matrix. As an approximation, we neglect the electromagnetic 
two-body production $T_{\pi\gamma}^{(NN)}$ and then the outgoing 
$\pi NN$ scattering state is then approximated by the free $\pi NN$ plane 
wave.

For the spin $( |s m\rangle )$ and isospin $( |t -\mu\rangle )$ part
of the two nucleon wave functions we use a coupled spin-isospin basis
$\ket{s m,\,t -\mu}$.  The antisymmetric final $NN$ plane wave function
thus has the form
\beq
  |\vec{p}_{1},\vec{p}_{2},s m,t -\mu \rangle =
  \frac{1}{\sqrt{2}}\left(
    |\vec{p}_{1}\rangle^{(1)}|\vec{p}_{2}\rangle^{(2)} - (-)^{s+t}
    |\vec{p}_{2}\rangle^{(1)}|\vec{p}_{1}\rangle^{(2)}\right)|s
  m\,,t -\mu\rangle\,,
\eeq
where the superscript indicates to which particle the ket refers.
In the case of charged pions, only the $t = 1$ channel contributes 
whereas for $\pi^{0}$ production both $t = 0$ and $t = 1$ channels
have to be taken into account. Then, the matrix element is given by 
\beqa
  {\mathcal M}_{sm m_{\gamma}m_d}^{(t\mu)}
  (\vec{k},\vec{q},\vec{p_1},\vec{p_2})&=& \langle
\vec{p}_{1},\vec{p}_{2},s m,t -\mu 
  |t_{\gamma\pi}^{NN}(\vec k,\vec q\,)|\vec{d}m_{d},00 \rangle \nonumber\\
&=&\frac{1}{2} \int
  \frac{d^{3}p^{\prime}_{1}}{(2\pi)^{3}} \int
  \frac{d^{3}p^{\prime}_{2}}{(2\pi)^{3}}
  \frac{M_{N}^2}{E^{\,\prime}_{1}E^{\,\prime}_2} 
\nonumber\\  && \times 
\sum_{m^{\prime}} \langle \,\vec{p}_{1}\vec{p}_{2},s
  m,t  -\mu |\, t_{\gamma\pi}^{NN}(\vec k,\vec q\,)
  |\,\vec{p}_{1}^{\,\prime}\vec{p}_{2}^{\,\prime}, 1 m^{\prime}, 0
  0 \rangle \nonumber \\
&& \times \langle\,\vec{p}_{1}^{\,\prime}\vec{p}_{2}^{\,\prime}, 1
  m^{\prime},00|\,\vec{d} m_{d},\,00\rangle
\eeqa
with 
\beqa
t_{\gamma\pi}^{NN}(\vec k,\vec q\,)=t_{\gamma\pi}^{N(1)}(\vec k,\vec q\,)
+t_{\gamma\pi}^{N(2)}(\vec k,\vec q\,)\,,\label{tmat-IA}
\eeqa
where $t_{\gamma\pi}^{N(j)}$ denotes the elementary production amplitude 
on nucleon ``$j$''.  The deuteron wave function has the form
\beq
  \langle\,\vec{p}_{1}\vec{p}_{2}, 1
  m,\,00|\,\vec{d} m_{d},00\rangle = (2\pi)^{3}
  \delta^{3}(\,
    \vec{d}-\vec{p}_{1}-\vec{p}_2 \,)
  \frac{\sqrt{2\,E_{1}E_{2}}}
  {M_{N}} \widetilde{\Psi}_{m,m_{d}}(\vec{p}_{NN})
\eeq
with
\beqa
  \widetilde{\Psi}_{m,m_{d}}(\vec{p}\,) =
  (2\pi)^{\frac{3}{2}}\sqrt{2E_{d}}
  \sum_{L=0,2}\sum_{m_{L}}i^{L}\,C^{L 1 1}_{m_{L} m m_{d}}\,
  u_{L}(p)Y_{Lm_{L}}(\hat{p}) \,.
\eeqa

Using (\ref{tmat-IA}) one finds in the laboratory 
system for the matrix element the following expression
\beqa
\label{tmat_IA_lab}
  {\mathcal M}_{sm m_{\gamma}m_d}^{(t\mu)}
  (\vec k,\vec q,\vec p_1,\vec p_2) &=&
 \sqrt{2}\sum_{m^{\prime}}\langle s 
  m,\,t -\mu|\,\Big( \langle
  \vec{p}_{1}|t_{\gamma\pi}^{N(1)}(\vec k,\vec q\,)|-\vec{p}_{2}\rangle
  \tilde{\Psi}_{m^{\prime},m_{d}}(\vec{p}_{2}) 
\nonumber\\  \hspace{1cm}& &
-(-)^{s+t}(\vec p_1 \leftrightarrow \vec p_2) 
\Big)\,|1 m^{\prime},\,00\rangle.
\eeqa
Note, that in (\ref{tmat_IA_lab}) the elementary production operator acts on nucleon
``1'' only. This matrix element possesses the obvious symmetry under the
interchange of the nucleon momenta
\beqa
  {\mathcal M}_{sm m_{\gamma}m_d}^{(t\mu)}
  (\vec k,\vec q,\vec p_2,\vec p_1) =(-)^{s+t+1}\,
  {\mathcal M}_{sm m_{\gamma}m_d}^{(t\mu)}
  (\vec k,\vec q,\vec p_1,\vec p_2) \,.
\eeqa

\section{Definition of a general polarization observable}
\label{sec5}
In this section the expressions for single-spin asymmetries are derived. 
For a given 
$\mathcal M$-matrix we compute the cross section for arbitrary polarized 
photons and initial deuterons by applying the density matrix formalism similar 
to that given by Arenh\"ovel~\cite{Aren88} for deuteron photodisintegration. 
The most general expression for all possible polarization observables in the 
reaction $d(\gamma,\pi)NN$ is given in terms of the transition 
$\mathcal{M}$-matrix by 
\beqa
\mathcal O &=& \sum_{\stackrel{smtm_{\gamma}m_d}{s^{\prime} 
m^{\prime}t^{\prime}m_{\gamma}^{\prime}m_d^{\prime}}}\int_{0}^{q_{\rm max}}dq
\int d\Omega_{p_{NN}}\mathcal{K}
\mathcal M^{(t^{\prime}\mu^{\prime})~\star}_{s^{\prime}m^{\prime},
m_{\gamma}^{\prime}m_d^{\prime}}\vec{\Omega}_{s^{\prime}m^{\prime}sm}
\mathcal M^{(t\mu)}_{sm,m_{\gamma}m_d}
\rho^{\gamma}_{m_{\gamma}m_{\gamma}^{\prime}} \rho^{d}_{m_dm_{d}^{\prime}}\,,
\nonumber \\ & &
\eeqa
where $\rho^{\gamma}_{m_{\gamma}m_{\gamma}^{\prime}}$ and
$\rho^{d}_{m_dm_{d}^{\prime}}$ denote the density matrices of initial
photon polarization and deuteron orientation, respectively, 
$\vec{\Omega}_{s^{\prime}m^{\prime}sm}$ is an operator associated with
the observable, which acts in the two-nucleon spin space and $\mathcal{K}$ 
is a phase space factor given in (\ref{rhos}). For further details we refer 
to~\cite{Aren88}.

As shown in~\cite{Aren88}, that all possible polarization observables for  
the reaction $\gamma d\to\pi NN$ with polarized photon beam and/or oriented 
deuteron target can be expressed in terms of the quantities
\beqa
\mathcal{V}_{IM} & = &
\frac{(-)^M\sqrt{2I+1}}{2\sqrt{15}}~\sum_{m_d^{\prime}m_d}~
(-)^{1-m_d^{\prime}}~
C^{112}_{m_dm_d^{\prime}M}~
\nonumber \\  & & \times 
\sum_{smtm_{\gamma}}\int_{0}^{q_{\rm max}}dq~\int d\Omega_{p_{NN}}~ \mathcal{K}~  
\mathcal M^{(t\mu)~\star}_{sm,m_{\gamma}m_d}~
\mathcal M^{(t\mu)}_{sm,m_{\gamma}m_d^{\prime}}\,,
\label{VIM}
\eeqa
and
\beqa
\mathcal{W}_{IM} & = &
\frac{(-)^M\sqrt{2I+1}}{2\sqrt{15}}~\sum_{m_d^{\prime}m_d}~
(-)^{1-m_d^{\prime}}~
C^{112}_{m_dm_d^{\prime}M}~
\nonumber \\  & & \times 
\sum_{smtm_{\gamma}}\int_{0}^{q_{\rm max}}dq~\int d\Omega_{p_{NN}}~ \mathcal{K}~  
\mathcal M^{(t\mu)~\star}_{sm,m_{\gamma}m_d}~
\mathcal M^{(t\mu)}_{s-m,m_{\gamma}-m_d^{\prime}}\,.
\eeqa

A quantity of great interest is the photon asymmetry $\Sigma$ for linearly 
polarized photons, which is defined as 
\beqa
\Sigma ~\frac{d^2\sigma}{d\Omega_{\pi}} & = & - \mathcal{W}_{00}\,.
\eeqa
It can be expressed in terms of the transition $\mathcal{M}$-matrix 
elements as follows
\beqa
\Sigma & = & \frac{2}{\mathcal F} ~\Re e \sum_{smtm_{d}}
\int_{0}^{q_{\rm max}}dq~\int d\Omega_{p_{NN}} ~\mathcal{K} 
~{\mathcal M}^{(t\mu)}_{sm +1m_d}
~{\mathcal M}^{(t\mu)~\star}_{sm -1m_d}\,,
\label{FSigma}
\eeqa
with
\beqa
\label{fff}
\mathcal F &=& \sum_{smtm_{\gamma}m_d}\int_{0}^{q_{\rm max}}dq \int d\Omega_{p_{NN}} 
  ~\mathcal{K}~ |{\mathcal M}^{(t\mu)}_{sm m_{\gamma}m_d}|^{2}\,.
\eeqa

The vector target asymmetry $T_{11}$ is defined as
\beqa
T_{11}~\frac{d^2\sigma}{d\Omega_{\pi}} & = & 2~\Im m \mathcal{V}_{11}\,.
\label{T11}
\eeqa
It can be expressed in terms of the transition $\mathcal{M}$-matrix as follows
\beqa
T_{11} & = & \frac{\sqrt{6}}{\mathcal F} ~\Im m\sum_{smtm_{\gamma}}
\int_{0}^{q_{\rm max}}dq~\int d\Omega_{p_{NN}}~ \mathcal{K}~
\Big[{\mathcal M}^{(t\mu)}_{smm_{\gamma}-1}
- {\mathcal M}^{(t\mu)}_{smm_{\gamma}+1}\Big]
{\mathcal M}^{(t\mu)~\star}_{smm_{\gamma}0}\,. 
\nonumber \\ & & 
\label{FT11}
\eeqa

The tensor target asymmetries $T_{2M}$ with $M=0,1,2$ are defined as 
\beqa
T_{2M} ~\frac{d^2\sigma}{d\Omega_{\pi}} & = & (2-\delta_{M0})~\Re e~ \mathcal{V}_{2M}\,,~~
  M=0,1,2\,.
\eeqa
These asymmetries can also be expressed in terms of the transition 
$\mathcal{M}$-matrix as follows
\beqa
T_{20} & = & \frac{1}{\sqrt{2}\mathcal F} \sum_{smtm_{\gamma}} 
\int_{0}^{q_{\rm max}}dq~\int d\Omega_{p_{NN}}~\mathcal{K}~\Big[~|{\mathcal M}^{(t\mu)}_{smm_{\gamma}+1}|^2 
+ |{\mathcal M}^{(t\mu)}_{smm_{\gamma}-1}|^2
\nonumber \\ & & \hspace{5.5cm}
- 2~ |{\mathcal M}^{(t\mu)}_{smm_{\gamma}0}|^2~\Big]\,,
\label{FT20}
\eeqa
\beqa
T_{21} & = & \frac{\sqrt{6}}{\mathcal F} ~\Re e\sum_{smtm_{\gamma}} 
\int_{0}^{q_{\rm max}}dq~\int d\Omega_{p_{NN}} ~\mathcal{K}~ \Big[{\mathcal M}^{(t\mu)}_{smm_{\gamma}-1} 
- {\mathcal M}^{(t\mu)}_{smm_{\gamma}+1}\Big] ~
{\mathcal M}^{(t\mu)~\star}_{smm_{\gamma}0}\,, 
\nonumber \\ & &
\label{FT21}
\eeqa
\beqa
T_{22} & = & \frac{2\sqrt{3}}{\mathcal F} ~\Re e\sum_{smtm_{\gamma}} 
\int_{0}^{q_{\rm max}}dq~\int d\Omega_{p_{NN}} ~\mathcal{K}~ {\mathcal M}^{(t\mu)}_{smm_{\gamma}-1} 
~{\mathcal M}^{(t\mu)~\star}_{smm_{\gamma}+1} \,. 
\label{FT22}
\eeqa

\section{Results and discussion}
\label{sec6}
Here we present and discuss the numerical results of the formalism developed
before in calculating the single-spin asymmetries of all the three isospin
channels of pion photoproduction on the deuteron in the $\Delta$(1232)-resonance
region. The contribution to the pion production amplitude in~(\ref{tmat_IA_lab}) 
is evaluated by taking the realistic $NN$ potential model for the deuteron
wave function. For our calculations we used the wave function of the Paris
potential~\cite{La+81}, which is in good agreement with $NN$ scattering
data~\cite{Dar03th}. We would like to remark, that as we see from the discussion 
below we have obtained essentially the same results for the linear photon and 
vector target asymmetries if we take the deuteron wave function of the Bonn r-space 
potential~\cite{MaH87} instead of the Paris one while small changes in the results are 
found for the tensor target asymmetries.

The discussion of our results is divided into three parts. The first part contains 
the discussion of the results for the photon asymmetry $\Sigma$ for linearly 
polarized photons as a function of pion angle $\theta_{\pi}$ in the laboratory frame 
at four different photon lab-energies. 
In the second part we discuss the results for the vector target asymmetry 
$T_{11}$. The results for the tensor 
target asymmetries $T_{20}$, $T_{21}$, and $T_{22}$ are discussed in the third part. 
In all parts, we give the calculations for all the three isospin channels of 
the reaction $d(\gamma,\pi)NN$. In order to investigate qualitatively the separate 
role of the contributions, the effect of various contributions of the single-nucleon operator, i.e., Born terms and 
$\Delta$(1232)-resonance term, is shown.

All the above mentioned single-spin asymmetries are calculated by integrating 
over the pion momentum $q$ and the polar angle $\theta_{p_{NN}}$ and the 
azimuthal angle $\phi_{p_{NN}}$ of the 
relative momentum $\vec p_{NN}$ of the two outgoing nucleons. These 
integrations are carried out numerically. The number of integration Gauss-points 
was being increased until the accuracy of calculated observable becomes good 
to 1$\%$.

\subsection{Photon asymmetry}
\label{sec61}
\begin{figure}[htb]
\includegraphics[scale=.75]{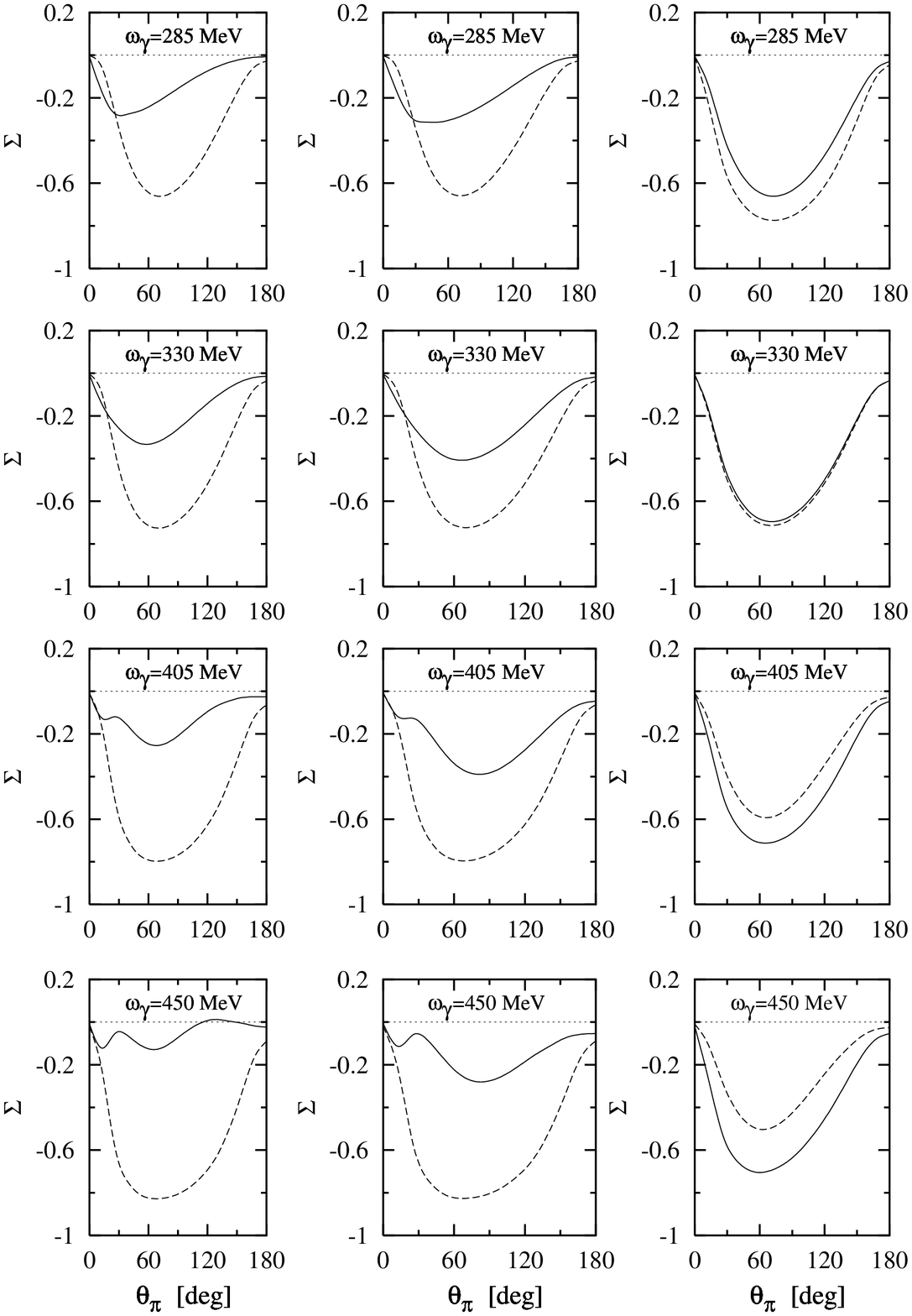}
\caption{Linear photon asymmetry $\Sigma$ of $\vec\gamma d \to\pi^-pp$ (left panels),
$\pi^+nn$ (middle panels), and $\pi^0np$ (right panels) as a
  function of pion angle $\theta_{\pi}$ in the laboratory frame at different
  photon lab-energies. Solid curves: full calculation; dashed curves:
  contribution of $\Delta$(1232)-resonance, i.e., without the Born terms.}
\label{sigma}
\end{figure}

Here we discuss our results for the photon asymmetry $\Sigma$ for 
linearly polarized photons for all the three different charge 
states of the pion of the reaction $d(\vec\gamma,\pi)NN$. The 
$\gamma$-asymmetry at four different values of the photon 
lab-energies are plotted in Fig.~\ref{sigma} for $\vec\gamma d\to \pi^-pp$ 
(left panels), $\pi^+nn$ (middle panels), and $\pi^0np$ (right panels) 
as a function of pion angle $\theta_{\pi}$ in the laboratory frame.
The solid curves show the results of the full calculation while the dotted 
ones show the contribution of the $\Delta$(1232)-resonance alone in order to 
clarify the importance of the Born terms. First of all, we see that the 
photon asymmetry has always a negative values at forward and backward 
emission pion angles for charged as well as for neutral pion channels. 
Only a very small positive value is found for $\pi^+$ production at 450 MeV. 
One notes qualitatively a similar behaviour for charged pion channels whereas 
a totally different behaviour is seen for the neutral pion channel.

For extreme forward and backward pion angles one sees, that the 
effect of Born contributions is relatively small in comparison to the 
results when $\theta_{\pi}$ changes from $60^{\circ}$ to $120^{\circ}$. 
One notices also, that the contribution from Born terms are much important in 
this region, in particular for charged pion channels. In the energy range of the 
$\Delta$(1232)-resonance, one sees that the contribution from Born terms are important in the case of charged 
pion channels. For the neutral pion channel we see, that this contribution is very small at 330 MeV. 
Since these calculations were done at 330 MeV, it is not surprising that the $\Delta$-contribution is dominant. 
For lower and higher energies, one sees again the sizeable effect from Born terms which arise from the Kroll-Rudermann term 
since it contributes only to the photoproduction of charged pions. One sees also, that $\Sigma$ is sensitive to the energy of the incoming photon.
 
We would like to remark here, that we have obtained essentially the same 
results for the linear photon asymmetry if we take the deuteron wave 
function of the Bonn r-space potential~\cite{MaH87} instead of the deuteron 
wave function of the Paris one~\cite{La+81}. We would also like to mention, 
that the curves in both cases are identical and therefore the results using 
the deuteron wave function of the Bonn potential are not shown in 
Fig.~\ref{sigma}. Finally, we observe that the interference of the Born terms 
with the $\Delta$(1232)-resonance contribution causes considerable changes in 
the linear photon asymmetry. Experimental measurements will give us more 
valuable information on this asymmetry.

\subsection{Vector target asymmetry}
\label{sec62}
\begin{figure}[htb]
\includegraphics[scale=.75]{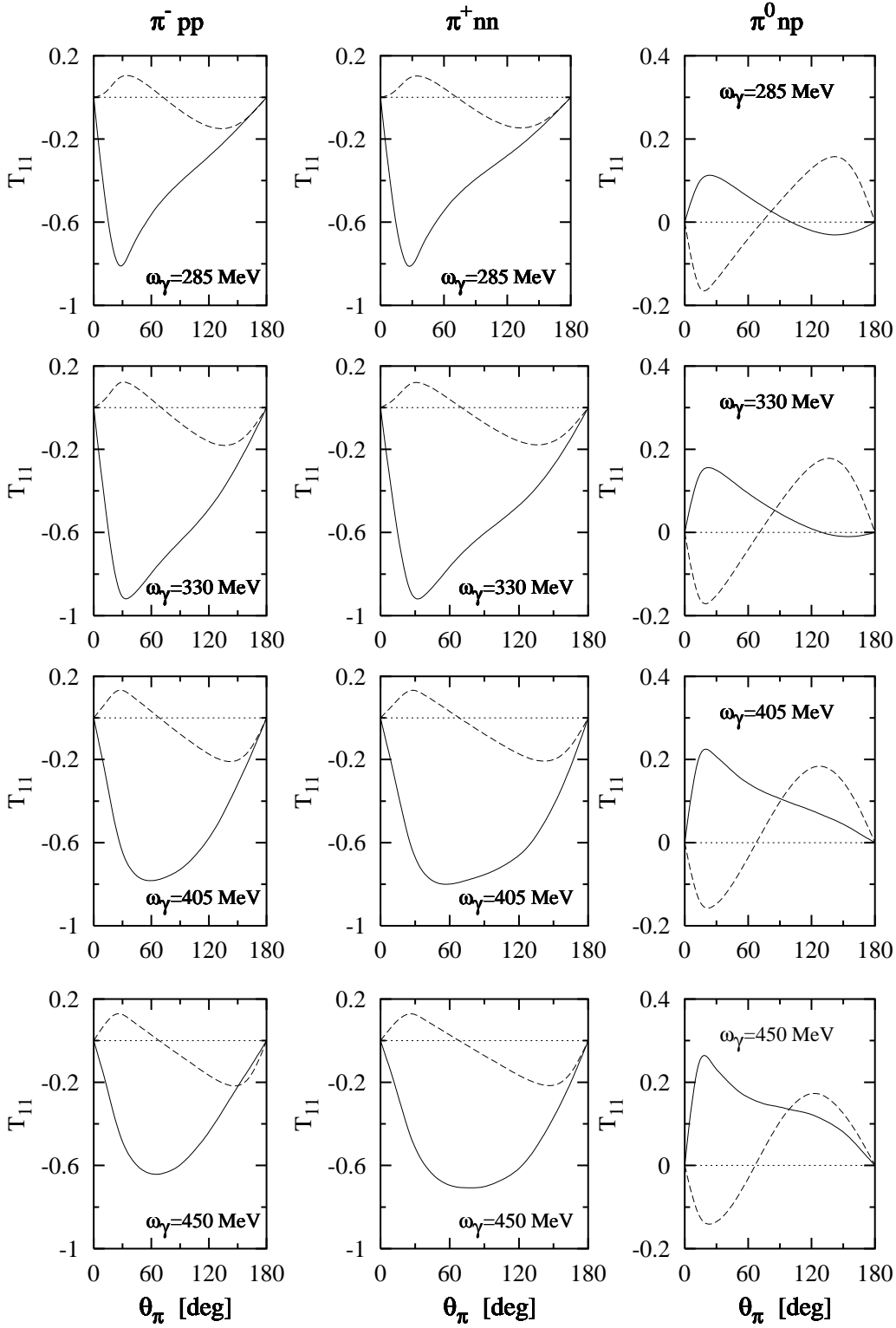}
\caption{Vector target asymmetry $T_{11}$ of $\gamma\vec d \to\pi^-pp$
(left panels), $\pi^+nn$ (middle panels), and $\pi^0np$ (right panels) as a
  function of pion angle $\theta_{\pi}$ in the laboratory frame at different
 photon lab-energies. 
  Notation of the curves as in Fig.~\ref{sigma}.}
\label{t11f}
\end{figure}

In this subsection we discuss our results for the vector target 
asymmetry $T_{11}$. Fig.~\ref{t11f} shows these results as a function of 
pion angle $\theta_{\pi}$ in the laboratory frame at four different values of 
photon lab-energies for $\gamma \vec d\to\pi^-pp$ (left panels), $\pi^+nn$ 
(middle panels), and $\pi^0np$ (right panels), respectively. The asymmetry 
$T_{11}$ clearly differs in size between charged and neutral pion 
photoproduction channels, being even opposite in phase. For charged pion 
photoproduction reactions we see from the left and middle panels of 
Fig.~\ref{t11f}, that the vector target asymmetry has always a negative 
values which mainly come from the Born terms. A small positive contribution 
from the $\Delta$-resonance is found only at pion forward angles. At backward 
angles, the negative values for $T_{11}$ come from an interference of the Born 
terms with the $\Delta$(1232)-resonance contribution. For all energies one 
observes at forward angle the strongest effect of the Born terms.

With respect to the neutral pion photoproduction channel, we see from the 
right panels of Fig.~\ref{t11f}, that the vector target 
asymmetry is always positive. For energies below the $\Delta$-resonance, 
a very small negative value is found at extreme backward pion angles while a 
relatively large positive value at forward angles is found. It is 
interesting to point out the importance of the Born terms in the charged pion 
production reactions in comparison to the contribution of the 
$\Delta$(1232)-resonance. This means, that $T_{11}$ is sensitive to the Born terms. 
The same effect was found by Blaazer {\it et al.}~\cite{Bla94} and Wilhelm and 
Arenh\"ovel~\cite{Wil95} for the coherent pion photoproduction reaction 
on the deuteron. The reason is that $T_{11}$ depends on the relative phase 
of the matrix elements as can be seen from (\ref{VIM}) and (\ref{T11}). 
It would vanish for a constant overall phase of the $t$-matrix, a case which 
is approximately realized if only the $\Delta$(1232)-amplitude is considered. 
Finally, we notice that $T_{11}$ is vanishes at $\theta_{\pi}=0$ and $\theta_{\pi}=\pi$ 
which is not the case for the linear photon asymmetry.

As discussed above in the case of photon asymmetry we obtained the same results for the vector target asymmetry 
if we take the deuteron wave function of the Bonn potential instead of the one of the Paris potential. Therefore, the results 
with the Bonn potential are not shown in Fig.~\ref{t11f}.

\subsection{Tensor target asymmetries}
\label{sec63}
\begin{figure}[htp]
\includegraphics[scale=0.75]{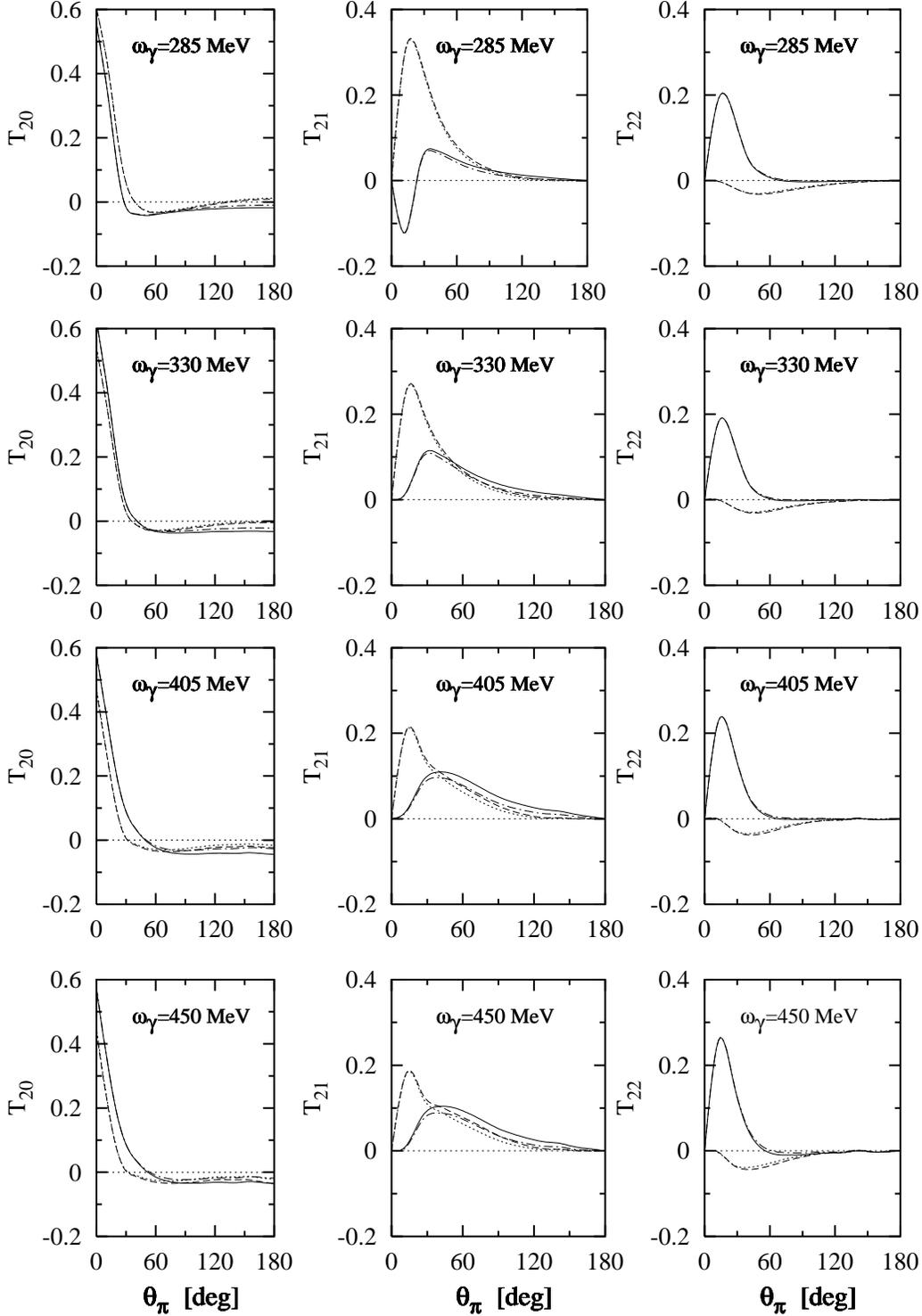}
\caption{Tensor target asymmetries $T_{20}$ (left panels), $T_{21}$ (middle 
  panels), and $T_{22}$ (right panels) of $\gamma\vec d\to \pi^-pp$ as a
  function of pion angle $\theta_{\pi}$ in the laboratory frame at different
  photon lab-energies. Solid (dash-dotted) curves: full calculation 
  using the deuteron wave function of Paris~\cite{La+81} (Bonn~\cite{MaH87})  
  potential model; dotted (long-dashed) curves: contribution of  
  $\Delta$(1232)-resonance using the deuteron wave function of 
  Paris~\cite{La+81} (Bonn~\cite{MaH87}) potential model.}
\label{tarpim}
\end{figure}

\begin{figure}[htp]
\includegraphics[scale=0.75]{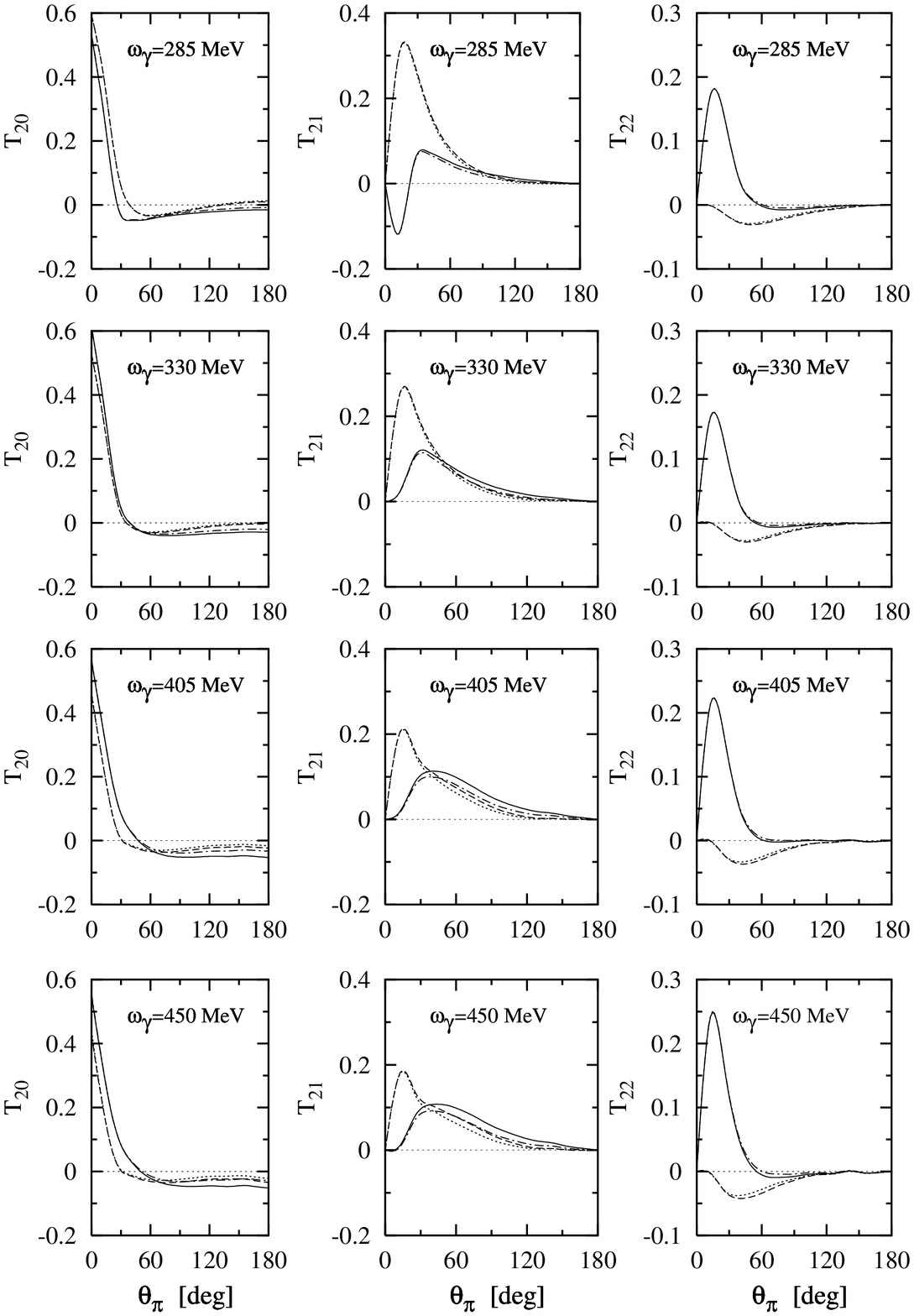}
\caption{Tensor target asymmetries $T_{20}$ (left panels), $T_{21}$ (middle 
  panels), and $T_{22}$ (right panels) of $\gamma\vec d\to \pi^+nn$
  as a function of pion angle $\theta_{\pi}$ in the laboratory frame at
  different photon lab-energies. Notation of the curves as in
  Fig.~\ref{tarpim}.}
\label{tarpip}
\end{figure}

\begin{figure}[htp]
\includegraphics[scale=0.75]{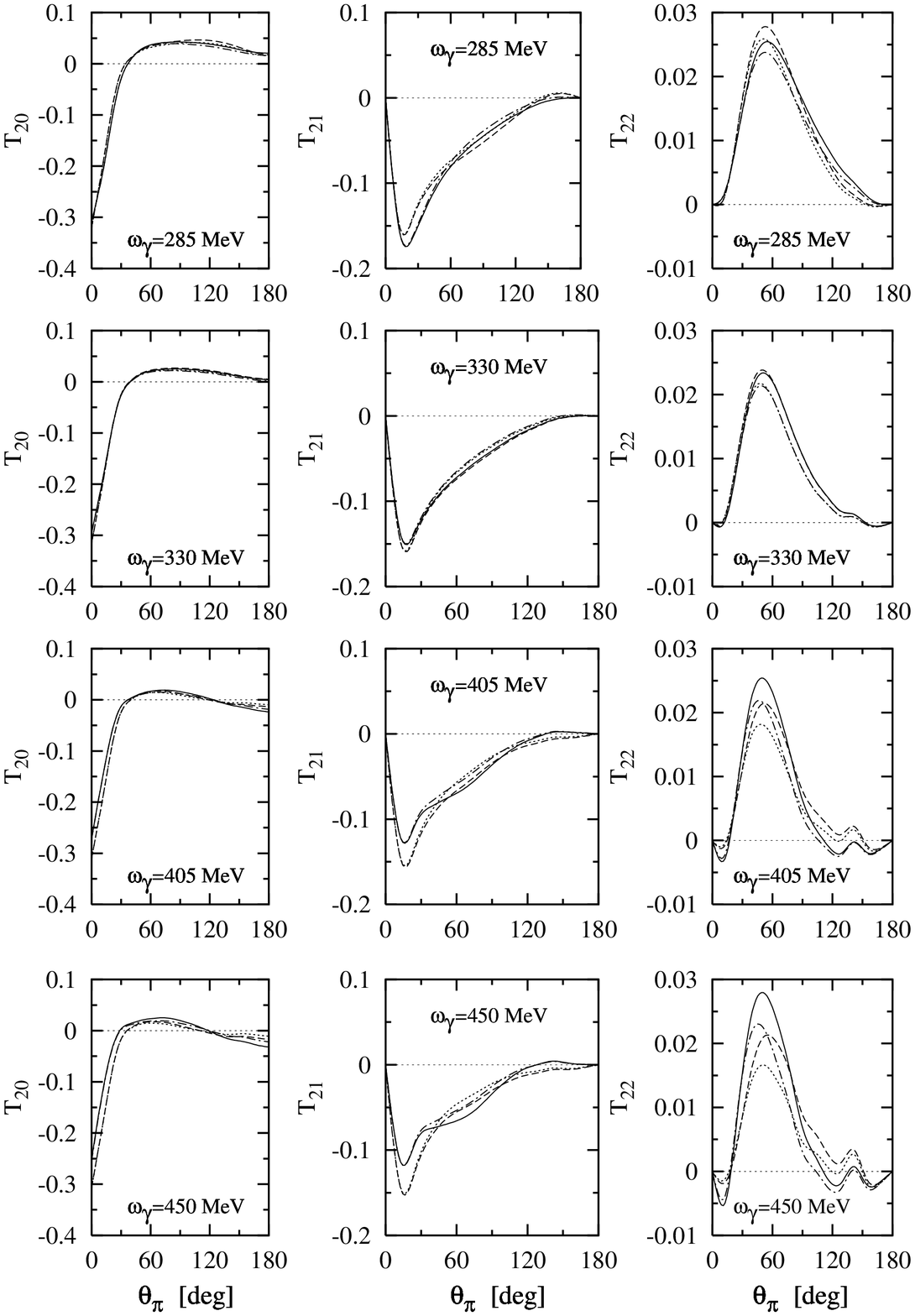}
\caption{Tensor target asymmetries $T_{20}$ (left panels), $T_{21}$ (middle 
  panels), and $T_{22}$ (right panels) of $\gamma\vec d\to \pi^0np$ as a 
  function of pion angle $\theta_{\pi}$
  in the laboratory frame at different photon lab-energies. Notation of the
  curves as in Fig.~\ref{tarpim}.}
\label{tarpiz}
\end{figure}

Let us present and discuss now the results of the tensor target asymmetries 
$T_{20}$, $T_{21}$, and $T_{22}$ as shown in Figs.~\ref{tarpim}, 
\ref{tarpip}, and \ref{tarpiz} for $\gamma \vec d\to\pi^-pp$, 
$\pi^+nn$, and $\pi^0np$, respectively. We start from the tensor asymmetry 
$T_{20}$ which is plotted in the left panels of Figs.~\ref{tarpim}, 
\ref{tarpip}, and \ref{tarpiz} for $\gamma \vec d\to\pi^-pp$, 
$\pi^+nn$, and $\pi^0np$, respectively, as a function of 
pion angle $\theta_{\pi}$ in the laboratory frame at four different values of 
photon lab-energies. The dotted curves represent the results for the 
contribution of the $\Delta$(1232)-resonance and the solid ones show the 
results when the Born terms are included. For $\gamma d\to\pi NN$ at forward 
and backward emission pion angles, the asymmetry $T_{20}$ 
allows one to draw specific conclusions about details of the reaction 
mechanism. In comparison to the results for photon and vector target asymmetries  
we found here, that the contribution from the Born terms is very small both for charged 
and neutral pion production channels. It is also noticeable, that for charged channels 
the asymmetry $T_{20}$ has a relatively large positive values at pion forward angles while a small negative ones at backward angles are found. 
For the neutral pion production channel we see, that $T_{20}$ has a negative values 
at forward angles and a positive ones at backward angles. Only for energies above the $\Delta$-resonance we note, that 
it has a small negative values at extreme backward angles.

The tensor target asymmetry $T_{21}$ of $\gamma \vec d\to\pi^-pp$, 
$\pi^+nn$, and $\pi^0np$ is plotted in the middle panels of 
Figs.~\ref{tarpim}, \ref{tarpip}, and \ref{tarpiz}, respectively. It is clear that 
$T_{21}$ differs in size between charged and neutral pion production channels. 
One notices, that for charged pion channels $T_{21}$ asymmetry is sensitive to the Born terms, in 
particular at forward pion angles. In the case of $\pi^0$ channel one sees, 
that the contribution of the Born terms is much less important at all energies. 
In comparison to the results for photon and vector target asymmetries  
we found also here, that the contribution from the Born terms is small both 
for charged and neutral pion production channels. It is also noticeable, 
that in the case of charged pion channels the asymmetry $T_{21}$ has a 
relatively large positive values at pion forward angles. 
For the neutral pion channel we see, that $T_{21}$ has a negative values 
at forward angles. Furthermore, as in the case of vector target asymmetry, we 
found that $T_{21}$ is vanishes at $\theta_{\pi}=0$ and $\theta_{\pi}=\pi$.

In the right panels of Figs.~\ref{tarpim}, \ref{tarpip}, and \ref{tarpiz} we depict our 
results for the tensor target asymmetry $T_{22}$ for the reactions 
$\gamma \vec d\to\pi^-pp$, $\pi^+nn$, and $\pi^0np$, respectively. 
One readily notes the importance of Born terms, in particular for charged 
pion channels at extreme forward pion angles. Like 
the results of the $T_{20}$ and $T_{21}$ asymmetries, the $T_{22}$ asymmetry 
is sensitive to the values of pion angle $\theta_{\pi}$. 
At $\theta_{\pi}=60^{\circ}$ we see, that the Born terms are 
important for $\pi^0$ production channel while these terms are very important 
for charged pion channels at extreme forward angles. Moreover, we found that 
$T_{22}$ is also vanishes at $\theta_{\pi}=0$ and $\theta_{\pi}=\pi$.

Finally, we would like to remark that we have obtained a different results 
for the tensor target asymmetries if we take the deuteron wave function of 
the Bonn potential instead of the one of the Paris potential. This difference 
is very clear in the case of neutral pion production channel as seen in the 
right panels of Fig.~\ref{tarpiz}.

\section{Conclusions}
\label{sec7}
In this paper we have studied incoherent single pion photoproduction 
on the deuteron in the first resonance region with special 
emphasis on single-spin asymmetries. For the elementary pion photoproduction
operator an effective Lagrangian model is used which is 
based on time-ordered perturbation theory and describes well the elementary
$\gamma N\to\pi N$ reaction. Particular attention was paid to
the single-spin asymmetries. We have 
presented results for the linear photon asymmetry $\Sigma$, vector target
asymmetry $T_{11}$ and tensor target asymmetries $T_{20}$, $T_{21}$, and
$T_{22}$. In particular, we have studied in detail the interference of the nonresonant 
background amplitudes with the dominant $\Delta$-excitation amplitude. The vector
target asymmetry $T_{11}$ has been found to be very sensitive to this interference. 
As already mentioned in the discussion above, interference of 
Born terms and the $\Delta$(1232)-contribution plays a significant role in the calculations. 
Unfortunately, there are no experimental data available to be compared to the 
spin observables we computed.

We would like to conclude that the results presented here for spin 
observables of $d(\gamma,\pi)NN$ can be used as a basis for the simulation
of the behaviour of polarization observables and for an optimal planning of
new polarization experiments of this reaction. It would be very interesting
to examine our predictions experimentally. 

As future refinements of the present model we consider the use of a more
sophisticated elementary production operator, which will allow one to extend
the present approach to higher energies. 
Future improvements should also include further investigations including
final state interaction as well as two-body effects.

\begin{ack}
I am gratefully acknowledge very useful discussions with Professor H.\
Arenh\"ovel as well as the members of his work group.
\end{ack}


\end{document}